\documentclass[twocolumn,aps,floatfix]{revtex4}
\usepackage{amssymb}
\usepackage{graphicx}

\mathchardef\ogon="012C%
\newcommand{\as}{a\kern-0.22em\lower.40ex\hbox{$_{\ogon}$}}
 
\begin{document}
 
\title{Ground state of two--component degenerate fermionic gases}
 
\author{Tomasz Karpiuk,$\,^1$ Miros{\l}aw Brewczyk,$\,^1$
        and Kazimierz Rz{\c a}\.zewski$\,^2$}
 
\affiliation{\mbox{$^1$ Uniwersytet w Bia{\l}ymstoku, ulica Lipowa 41,
                        15-424 Bia{\l}ystok, Poland}  \\
\mbox{$^2$ Centrum Fizyki Teoretycznej PAN, Aleja Lotnik\'ow 32/46, 
                        02-668 Warsaw, Poland}  }

\date{\today} 
 
\begin{abstract}
We analyze the ground state of the two--component gas of trapped ultracold
fermionic atoms. We neglect the forces between atoms in the same hyperfine
state (the same component). For the case when the forces between distinguishable 
atoms (i.e., atoms in different hyperfine states) are repulsive (positive mutual 
scattering length), we find the existence of critical interaction strength above 
which one atomic fraction expels the other from the center of the trap. When 
atoms from different components attract each other (negative mutual scattering 
length) the ground state of the system dramatically changes its structure for 
strong enough attraction -- the Cooper pairs built of atoms in different 
hyperfine states appear.
 
PACS Number(s):05.30.Fk, 03.75.Ss
\end{abstract}

\maketitle

\section{Introduction}
After the first experimental achievement of quantum degeneracy in potassium 
$^{40}$K gas \cite{Jin} followed by the cooling of other elements ($^6$Li) 
below the Fermi temperature \cite{Li}, the interest in ultracold atomic Fermi 
systems has increased significantly. Since at very low temperatures the s--wave
scattering in a single--component spin--polarized Fermi gas is excluded by the
statistics, the experimental realization of quantum degeneracy requires actually
trapping of two kinds of atoms. In fact, in Ref. \cite{Jin} the simultaneous
trapping of different spin states of $^{40}$K is reported as a way which
enables cooling the fermions by rethermalizing collisions undergoing between
atoms in different spin states. Another idea is to employ the sympathetic cooling
technique in a mixture of fermionic and bosonic isotopes like $^6$Li and
$^7$Li \cite{Li} or even in two--species Fermi--Bose mixtures like
$^6$Li -- $^{23}$Na \cite{Ketterle} or $^{40}$K -- $^{87}$Rb \cite{Inguscio}.

Several aspects of the physics of degenerate Fermi gases have been investigated 
in the past few years. In Ref. \cite{Butts} the properties of a harmonically
trapped spin--polarized (i.e., non--interacting) Fermi gas is studied by using 
the Thomas--Fermi approximation. This work was extended in several ways. One 
possibility is to consider the dipole--dipole forces as in Ref. \cite{Krzysiek},
where the ground state of fermionic dipoles in the normal phase is analyzed and 
mechanical instabilities for large enough number of particles or the dipole moment
are discovered. Another way is going beyond the semiclassical approximation by
including the discrete nature of the trap levels as well as by introducing the
second component and the interaction between atoms from different components 
\cite{Bruun}.

Not only static but also dynamic properties of cold Fermi gases have been
investigated. For example, the collective excitations in degenerate fermionic 
gas are discussed within the hydrodynamic approximation in Ref. \cite{Bruun1}
and based on the sum rules in Ref. \cite{Stringari}. There has been developed 
an idea of generating solitons and vortices in a one--component Fermi gas 
in a normal phase by using the phase imprinting technique \cite{Tomek}. The
presence of vorticity in a trapped gas rotating at low angular velocity is
analyzed in Ref. \cite{Stringari1}.

Much emphasis has been also put on the achievement of superfluid phase transition.
There was a proposal to reach the Cooper--paired state at temperatures comparable
to the Fermi energy via a Feshbach resonance \cite{Holland} for a short--range
interaction between distinguishable fermions. It is plausible that the fermionic
atoms confined in an optical lattice could also undergo the phase transition at
high temperatures (one tenth of the free--space Fermi energy) independently of
the scattering length \cite{HTc}. The possibility of p--wave pairing through
the Feshbach resonance for a short--range interaction in a single--component
Fermi gas was discussed in \cite{Bohn} whereas the achievement of the superfluid
transition in dipolar gases (again in a one--component case) in Ref. \cite{Baranov}.

Structure of binary Bose--Einstein condensates have been studied extensively
over the last years \cite{BEC,Marek}. In Ref. \cite{Marek} possible classes 
of solutions within Thomas--Fermi approximation are given and verified 
by numerical integration of the coupled Gross--Pitaevskii equations. For the 
clarity only the case when all atoms repel each other is considered. It was 
found that for some range of parameters (scattering lengths) the ground state 
of binary system is a symmetry--breaking solution. Moreover, it turns necessary 
to go beyond the Thomas--Fermi approximation to obtain asymmetric ground state 
because the contribution of the kinetic energy (neglected within Thomas--Fermi 
approximation) is substantial in this case.

The similar approach but for a two--component Fermi gas is reported in Ref.
\cite{Roth} where the role of s-- and p--wave interactions is investigated
within the Thomas--Fermi approximation. The existence of the critical particle 
number was shown for pure repulsive s--wave interaction above which the spatial 
separation of both components appears. However, after the separation is reached, 
the Thomas--Fermi approach does not yield the unique answer. Further assumptions 
are necessary. For example, consideration of configurations with minimal interface 
as energetically favorable leads to asymmetric ground state of two--component Fermi 
gas. For attractive s--wave interaction Ref. \cite{Roth} predicts the collapse of 
the system for large enough number of atoms similarly to the case of Bose--Einstein 
condensate.

In this paper we analyze the ground state of two--component trapped ultracold
Fermi gas and demonstrate the existence of several characteristic coupling 
strengths. To this end, we use the mean--field Hartree--Fock method and also BCS
related approach when necessary (in the case when different kinds of atoms
attract each other). So, opposite to Ref. \cite{Roth}, we go beyond the Thomas--Fermi 
approximation. On the other hand, we are restricted by numerics to rather small 
sample of atoms. Moreover, in this paper we present results of only one--dimensional
calculations.

The paper is organized as follows. In Sec. \ref{secequ}, based on the Lagrange 
formalism, we derive the equations which govern the two--component gas of 
trapped ultracold fermionic atoms. Sec. \ref{secdensity} defines the mixed 
two--particle density matrix and calculates densities of each component. In 
Secs. \ref{plus} and \ref{minus} we discus the properties of the ground state 
of the system within the Hartree--Fock and BCS approaches. Finally, we conclude 
in Sec. \ref{concl}.

\section{Equations describing two--component Fermi gas at zero
         temperature}
\label{secequ}

The time--dependent many--body Schr\"odinger equation can be derived as 
the Euler--Lagrange equation from the following Lagrangian density
\begin{eqnarray}
&&\mathcal{L} = \frac{i\hbar}{2} \Psi^* \frac{\partial \Psi}{\partial t}
- \frac{i\hbar}{2} \Psi \frac{\partial \Psi^*}{\partial t}
- \frac{\hbar^2}{2 m} \sum_{i=1}^N \nabla_i \Psi^* \nabla_i  \Psi
\nonumber  \\
&&- \sum_{i=1}^N V_{trap}({\bf r}_i) \Psi^* \Psi 
- \sum_{i<j} V_{int}({\bf r}_i-{\bf r}_j) \Psi^* \Psi
\label{Lagden}
\end{eqnarray}
considered as a function of the many--body wave function 
$\Psi ({\bf r}_1,...,{\bf r}_N)$ and its spatial and time
derivatives. At low temperatures both components can be treated as
non--interacting Fermi gases since the s--wave scattering is absent for
spin--polarized fermions. The only interaction left is the repulsion
or attraction between atoms of different spins. At zero temperature 
the wave function of two--component ($N + N$ atoms) Fermi system is 
assumed to be the product of two Slater determinants
\begin{eqnarray}
&&\Psi ({\bf x}_1,...,{\bf x}_N;{\bf y}_1,...,{\bf y}_N) =  \nonumber \\
&& \nonumber \\
&&\frac{1}{\sqrt{N!}} \left |
\begin{array}{lllll}
\varphi_1^{(1)}({\bf x}_1) & . & . & . & \varphi_1^{(1)}({\bf x}_N) \\
\phantom{aa}. &  &  &  & \phantom{aa}. \\
\phantom{aa}. &  &  &  & \phantom{aa}. \\
\phantom{aa}. &  &  &  & \phantom{aa}. \\
\varphi_N^{(1)}({\bf x}_1) & . & . & . & \varphi_N^{(1)}({\bf x}_N)
\end{array}
\right |    \times    \nonumber \\
&& \nonumber \\
&&\frac{1}{\sqrt{N!}} \left |
\begin{array}{lllll}
\varphi_1^{(2)}({\bf y}_1) & . & . & . & \varphi_1^{(2)}({\bf y}_N) \\
\phantom{aa}. &  &  &  & \phantom{aa}. \\
\phantom{aa}. &  &  &  & \phantom{aa}. \\
\phantom{aa}. &  &  &  & \phantom{aa}. \\
\varphi_N^{(2)}({\bf y}_1) & . & . & . & \varphi_N^{(2)}({\bf y}_N)
\end{array}
\right |  \, . 
\label{Slater}
\end{eqnarray}

Inserting the wave function (\ref{Slater}) into the Lagrangian obtained
based on the Lagrangian density (\ref{Lagden}) and integrating each term 
over the appropriate $(N-1)$ coordinates for each component one obtains 
the effective Lagrangian density
\begin{eqnarray}
&&\mathcal{L'} = \frac{i\hbar}{2} \sum_{i=1}^N \varphi_i^{(1) *}
\frac{\partial \varphi_i^{(1)}}{\partial t} -
\frac{i\hbar}{2} \sum_{i=1}^N \varphi_i^{(1)}
\frac{\partial \varphi_i^{(1) *}}{\partial t}  \nonumber  \\
&&+ \frac{i\hbar}{2} \sum_{i=1}^N \varphi_i^{(2) *} 
\frac{\partial \varphi_i^{(2)}}{\partial t} -
\frac{i\hbar}{2} \sum_{i=1}^N \varphi_i^{(2)}
 \frac{\partial \varphi_i^{(2) *}}{\partial t}   \nonumber  \\
&&- \frac{\hbar^2}{2 m} \sum_{i=1}^N \nabla \varphi_i^{(1) *}
\nabla \varphi_i^{(1)}
- \frac{\hbar^2}{2 m} \sum_{i=1}^N \nabla \varphi_i^{(2) *}
\nabla \varphi_i^{(2)}   \nonumber  \\
&&- V_{trap}^{(1)} \sum_{i=1}^N  | \varphi_i^{(1)} |^2
- V_{trap}^{(2)} \sum_{i=1}^N  | \varphi_i^{(2)} |^2  \nonumber  \\
&&-g \sum_{i=1}^N | \varphi_i^{(1)} |^2  
   \sum_{i=1}^N | \varphi_i^{(2)} |^2 
\label{effective}
\end{eqnarray}
which depends on one--particle orbitals of both types and their spatial and 
time derivatives. Deriving Eq. (\ref{effective}) we have assumed the
contact interaction between distinguishable fermions with $g$ being the
coupling constant. Like in experiment \cite{Jin} both atomic fractions move
in harmonic potentials and the ratio of frequencies is equal to 
$\omega^{(1)}/\omega^{(2)}=\sqrt{9/7}$. Time--independent version of 
the Euler--Lagrange equations originating from the effective Lagrangian 
density (\ref{effective}) is given by
\begin{eqnarray}
&&-\frac{\hbar^2}{2 m} \nabla^2 \varphi_i^{(1)} + V_{trap}^{(1)} \varphi_i^{(1)}
+ g \sum_{j=1}^N |\varphi_j^{(2)}|^2 \varphi_i^{(1)} = \varepsilon_i^{(1)}
\varphi_i^{(1)}  \nonumber  \\
&&-\frac{\hbar^2}{2 m} \nabla^2 \varphi_i^{(2)} + V_{trap}^{(2)} \varphi_i^{(2)}
+ g \sum_{j=1}^N |\varphi_j^{(1)}|^2 \varphi_i^{(2)} = \varepsilon_i^{(2)}
\varphi_i^{(2)}  \nonumber  \\
&&i=1,2,...,N
\label{HFeq}
\end{eqnarray}
with $\varepsilon_i^{(1)}$ and $\varepsilon_i^{(2)}$ being the single--particle
Hartree--Fock energies. The total energy of the system is expressed in the 
form
\begin{eqnarray}
&&<\Psi|{\rm H}|\Psi> = \sum_{i=1}^N \varepsilon_i^{(1)}  +
\sum_{i=1}^N \varepsilon_i^{(2)}   \nonumber \\
&&- g \sum_{i,j=1}^N \int \left| \varphi_i^{(1)}({\bf x}) \right|^2
\left| \varphi_j^{(2)}({\bf y}) \right|^2 d {\bf x} \: d {\bf y}  \, .
\label{totenergy}
\end{eqnarray}
We solve numerically the set of Eqs. (\ref{HFeq}) to find the
ground state of two--component degenerate fermionic gases.

\section{One-- and two--particle density matrices}
\label{secdensity}

Mixed two--particle density function is defined as
\begin{eqnarray}
&&\rho_2^{(1)(2)} ({\bf x}_1,{\bf y}_1;{\bf x}'_{\!1},{\bf y}'_{\!1}) = 
\nonumber   \\
&&< \hat{\psi}^{(1)\dagger}({\bf x}_1)   
    \hat{\psi}^{(2)\dagger}({\bf y}_1)
    \hat{\psi}^{(1)}({\bf x}'_{\!1})
    \hat{\psi}^{(2)}({\bf y}'_{\!1})  >
\label{denmat}
\end{eqnarray}
or equivalently in the position representation
\begin{eqnarray}
&&\rho_2^{(1)(2)} ({\bf x}_1,{\bf y}_1;{\bf x}'_{\!1},{\bf y}'_{\!1}) = 
\nonumber   \\\     
&&\int  
\Psi ({\bf x}_1,{\bf x}_2,...,{\bf x}_{\!N};
{\bf y}_1,{\bf y}_2,...,{\bf y}_{\!N})    \nonumber  \\ 
&&\times \Psi^* ({\bf x}'_{\!1},{\bf x}_2,...,{\bf x}_{\!N};
{\bf y}'_{\!1},{\bf y}_2,...,{\bf y}_{\!N})   \nonumber  \\
&&\times d {\bf x}_2 ... d {\bf x}_{\!N} \; d {\bf y}_2 ... d {\bf y}_{\!N}
\label{denmat1}
\end{eqnarray}
One--particle densities are derived from mixed two--particle density functions 
(\ref{denmat},\ref{denmat1}) by contracting over $\bf x$ or $\bf y$ variables
\begin{eqnarray}
&&\rho_1^{(1)} ({\bf x}_1;{\bf x}'_{\!1}) =
  < \hat{\psi}^{(1)\dagger}({\bf x}_1)   
    \hat{\psi}^{(1)}({\bf x}'_{\!1}) >   \nonumber   \\
&&\rho_1^{(2)} ({\bf y}_1;{\bf y}'_{\!1}) =
  < \hat{\psi}^{(2)\dagger}({\bf y}_1)
    \hat{\psi}^{(2)}({\bf y}'_{\!1})  >
\label{mg1}
\end{eqnarray}
or equivalently
\begin{eqnarray}
&&\rho_1^{(1)} ({\bf x}_1;{\bf x}'_{\!1}) = \int 
\rho_2^{(1)(2)} ({\bf x}_1,{\bf y}_1;{\bf x}'_{\!1},{\bf y}_{1}) 
\; d {\bf y}_1    \nonumber \\
&&\rho_1^{(2)} ({\bf y}_1;{\bf y}'_{\!1}) = \int
\rho_2^{(1)(2)} ({\bf x}_1,{\bf y}_1;{\bf x}_{1},{\bf y}'_{\!1})
\; d {\bf x}_1
\end{eqnarray}

In the Hartree--Fock state given by (\ref{Slater}) two--particle density 
matrix, since single--particle orbitals are orthogonal, reduces to
\begin{eqnarray}
&&\rho_2^{(1)(2)} ({\bf x}_1,{\bf y}_1;{\bf x}'_{\!1},{\bf y}'_{\!1}) = 
\nonumber   \\\     
&&\frac{1}{N^2} \sum_{i,j}^N \varphi_i^{(1)*}({\bf x}_1)
\varphi_j^{(2)*}({\bf y}_1) \varphi_i^{(1)}({\bf x}'_{\!1})
\varphi_j^{(2)}({\bf y}'_{\!1})
\end{eqnarray}
and is always a product of one--particle density matrices (two--particle
correlation function vanishes)
\begin{eqnarray}
&&\rho_1^{(1)} ({\bf x}_1;{\bf x}'_{\!1}) = \frac{1}{N}
\sum_{i}^N \varphi_i^{(1)*}({\bf x}_1) \varphi_i^{(1)}({\bf x}'_{\!1})
\nonumber  \\
&&\rho_1^{(2)} ({\bf y}_1;{\bf y}'_{\!1}) = \frac{1}{N}
\sum_{i}^N \varphi_i^{(2)*}({\bf y}_1)  \varphi_i^{(2)}({\bf y}'_{\!1})
\end{eqnarray}
The density of each component equals the diagonal part of corresponding
one--particle density matrix and for the Hartree--Fock state (\ref{Slater})
is given by
\begin{eqnarray}
&&\rho^{(1)} ({\bf x}_1) = \frac{1}{N}  
\sum_i^N |\varphi_i^{(1)}({\bf x}_1)|^2
\nonumber \\
&&\rho^{(2)} ({\bf y}_1) = \frac{1}{N}
\sum_i^N |\varphi_i^{(2)}({\bf y}_1)|^2
\end{eqnarray}
In Section \ref{minus} we show that for strong enough attraction
between atoms better description of the ground state is BCS like
one with significant correlations developed.

\section{Positive scattering length}
\label{plus}

In this Section we consider the ground state of one--dimensional
two--component Fermi system with repulsive forces between atoms of 
different components. We solve numerically the set of Eqs. (\ref{HFeq}) 
(in its one--dimensional version) by using iterative procedure. At zero step 
we pick up the Hartree--Fock orbitals, for example as the eigenstates of 
the pure harmonic potentials for each component -- these functions are
called zero--step orbitals. Then we calculate potentials 
which come from the presence of other component and build 
Hamiltonian matrices for both parts of the system. Matrix elements are 
calculated in sets of basis functions which include corresponding 
harmonic oscillator eigenstates. The number of basis functions is typically 
much bigger than the number of particles $N$. At each step both matrices 
are diagonalized and new single--particle wave functions are obtained as 
well as single--particle energies. This procedure is continued until all 
quantities remain unchanged with prescribed accuracy.

\begin{figure}[hbt]
\resizebox{2.9in}{2.3in}
{\includegraphics{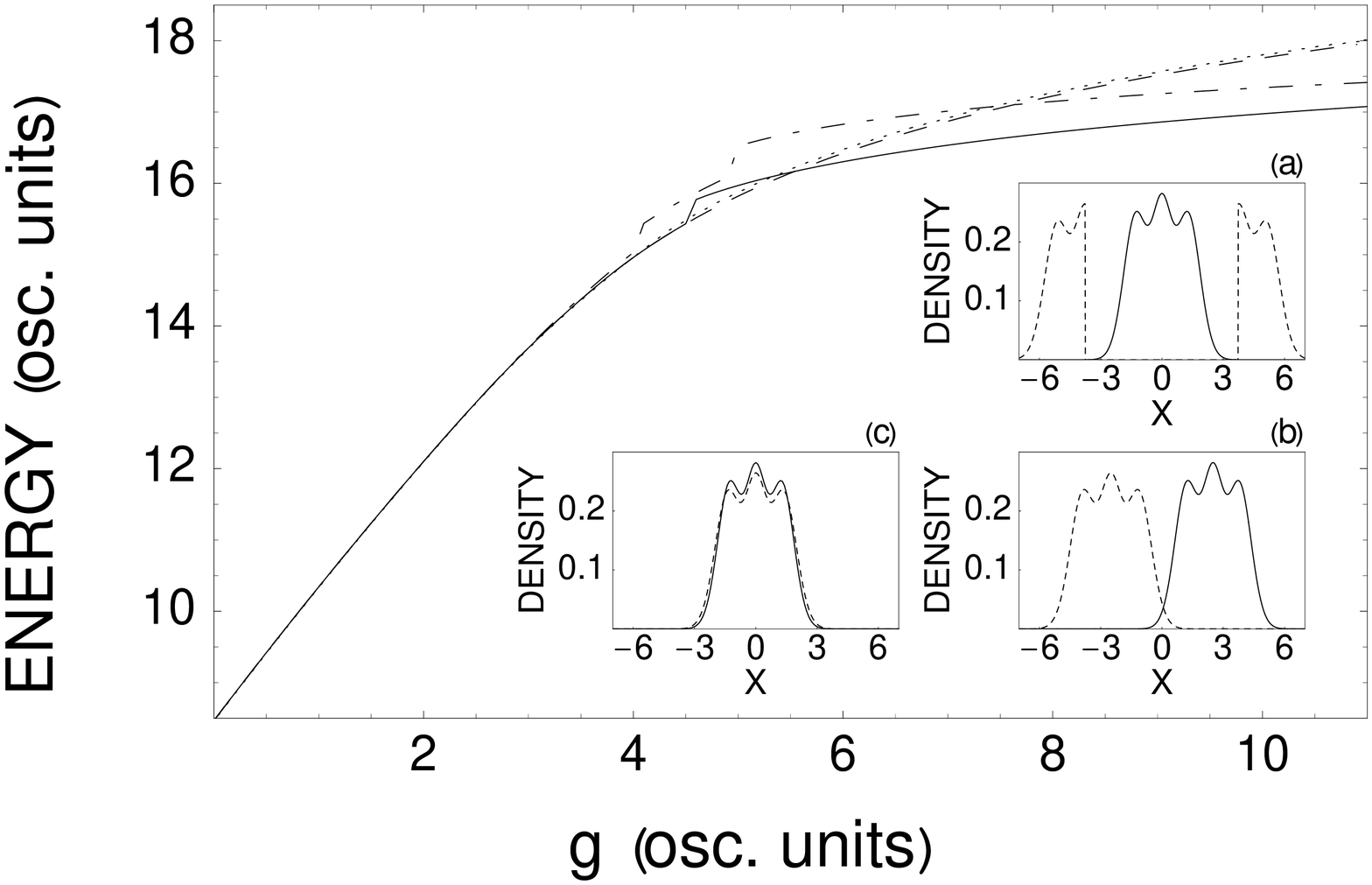}}
\caption{
Energy of $3 + 3$ atoms system as a function of positive interaction
strength obtained within the Hartree--Fock approach. For weak enough
repulsion (g $< 4$) all initial conditions lead to the same density 
profile which is the density of the ground state. For strong repulsion 
the density profile of the ground state changes qualitatively its shape 
(see Fig. \ref{denpos}). 
Various sets of zero--step orbitals give different local minima of energy.
Insets show the zero--step densities used in the numerics: (a) gives
the results marked by solid line, (b) corresponds to dotted--dashed line,
and (c) leads to both dotted and dashed lines. 
See text for further description of the procedure.}
\label{gplus}
\end{figure}

\begin{figure}[htb]
\resizebox{3.2in}{4.in}
{\includegraphics{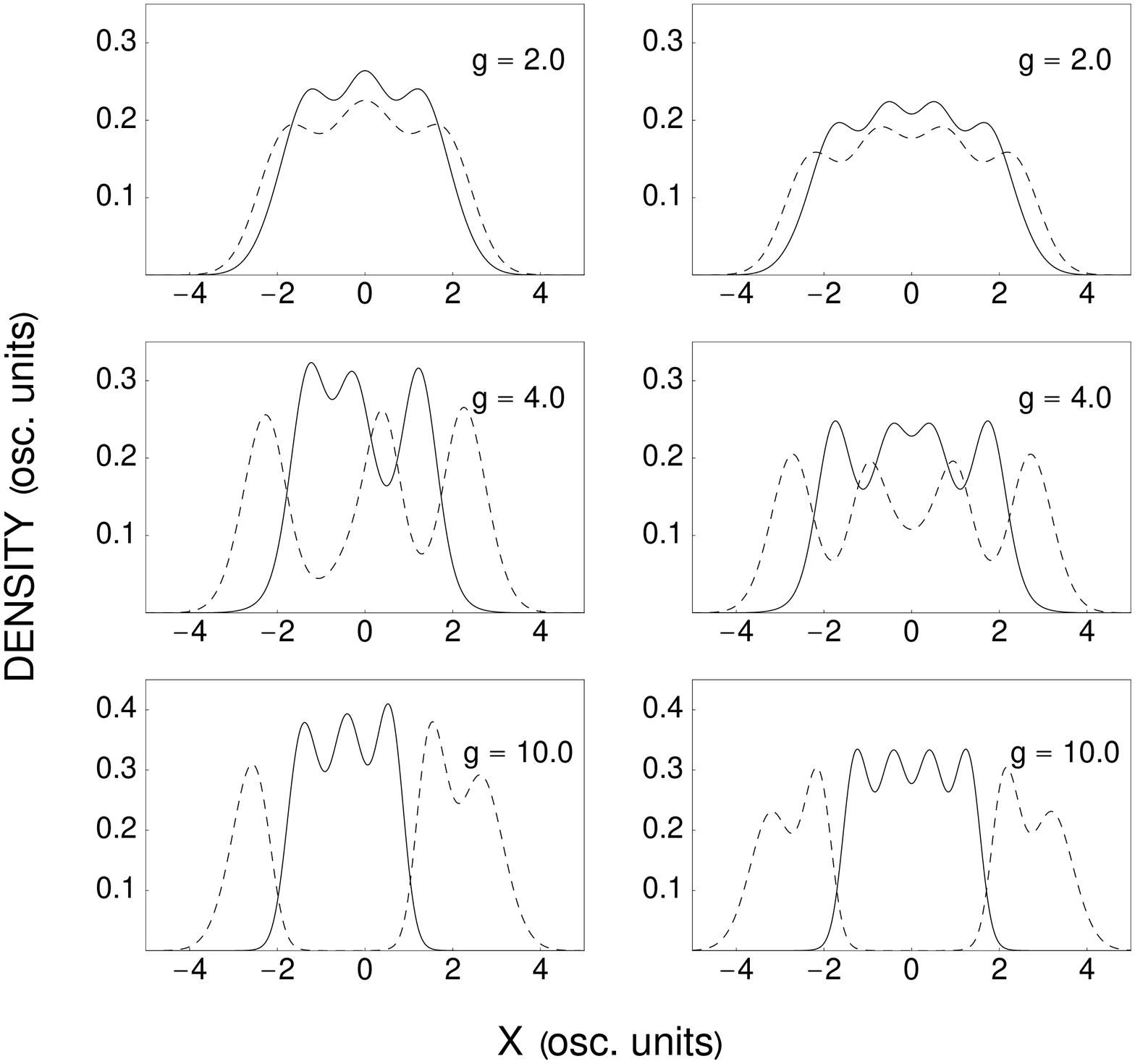}}
\caption{Density profiles of the ground state of two--component Fermi
gas for various positive interaction strengths. Left and right columns
show the case of $3+3$ and $4+4$ atoms systems, respectively. For small 
coupling constant ($g < 4$) both fractions almost fully overlap each other 
whereas beyond the short intermediate interval (around $g=4$) fractions 
separate. Full curves correspond to the first (higher frequency trap) 
component. The broken curves indicate the second component confined in a
lower frequency trap.}
\label{denpos}
\end{figure}

\begin{figure}[hbt]
\resizebox{2.7in}{2.1in}
{\includegraphics{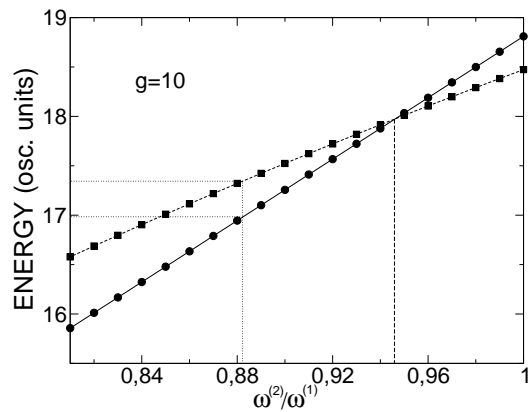}}
\caption{The energy of an asymmetric density profile states (boxes) and
symmetric ones (circles) as a function of the ratio
$\omega^{(2)}/\omega^{(1)}$ for $3+3$ atoms system. 
The vertical dotted line shows the geometry used
in this paper.}
\label{en}
\end{figure}

\begin{figure}[htb]
\resizebox{2.5in}{4.4in}
{\includegraphics{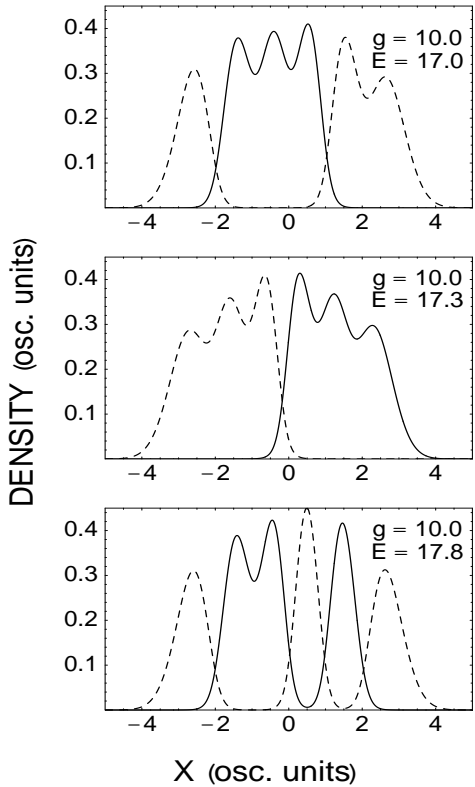}}
\caption{Density profiles for the coupling constant $g=10$. The upper frame 
corresponds to the ground state whereas the middle and lower ones to the
excited states which show qualitatively different patterns; separation
(the middle one) and the domain structure (the lowest one).}
\label{denpos10}
\end{figure}

We summarize our results in Fig. \ref{gplus}. There exists the critical
value of the interaction strength $(g \approx 4$  for $3+3$ atoms system) 
above which the density profile of the system changes qualitatively its shape 
(see Fig. \ref{denpos}). It is profitable from energetic point of view that the 
component confined in a lower frequency trap is repelled from the center of the 
trap. Although the potential energy is increased in such a way, the total energy, 
however, is decreased because the interaction (repulsion) is weakened. The solid 
line in Fig. \ref{gplus} shows the energy of the ground state of two--component 
system. Its density is presented in Fig. \ref{denpos} (left column). 
The fact that the density
is not symmetric with respect to the center of the trap (the lowest frame) is 
explained by the presence of odd number of atoms in each fraction. For example, 
for $4+4$ atoms system the density of the ground state is symmetric (see the right
column of frames in Fig. \ref{denpos}). The interplay
between the potential and the interaction energies of the gases is even more
subtle than just described. It turns out that the symmetry of the ground state 
strongly depends on the geometry of the traps as shown in Fig. \ref{en} for a
particular value of the coupling strength. The vertical dotted line indicates the 
ratio of the trap frequencies used throughout this paper. Depending on this ratio
the ground state of the system switches between the asymmetric density profile
(boxes in Fig. \ref{en}) like in the middle frame in Fig. \ref{denpos10} and 
symmetric one (circles in Fig. \ref{en})) as in the upper frame in Fig. 
\ref{denpos10}. The vertical 
dashed line marks the geometry ($\omega^{(2)}/\omega^{(1)}$=0.945) in which
the ground state of the system becomes degenerate. We have checked that 
increasing the repulsion strength between the components the ratio
$\omega^{(2)}/\omega^{(1)}$ that favors the degeneracy of the ground state
is getting lower. Eventually, for $g > 44$ the ground state of $3+3$ atoms 
system becomes asymmetric under the considered trapping potentials. 

\begin{figure}[htb]
\resizebox{2.5in}{4.4in}
{\includegraphics{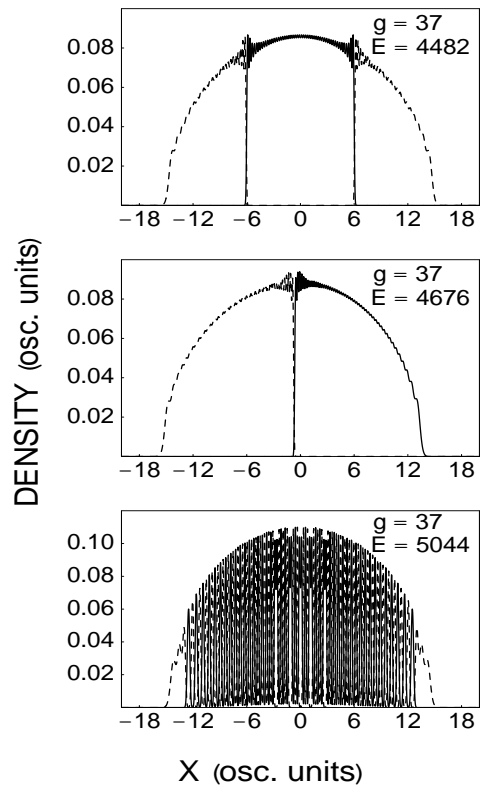}}
\caption{The sequence of frames similar to that in Fig. \ref{denpos10} but
for the system of $50 + 50$ atoms and  $g=37$.
Solid curves show the densities of the first component (loaded into higher
frequency trap) whereas the dashed ones correspond to the second component
(loaded into lower frequency trap).
The lowest frame is, in fact, the case of very dense domain structure.}
\label{denpos50}
\end{figure}

For large enough $g$ we find several local energy minima depending on the 
chosen set of zero--step orbitals. The lowest minimum is accessible provided
one uses for the second component harmonic oscillator eigenstates which are 
first cut at the center and then pulled out as a zero--step orbitals 
(inset (a) in Fig. \ref{gplus}). If plain harmonic oscillator eigenstates are 
taken as the initial wave functions (inset (c) in Fig. \ref{gplus})
one follows the dashed
line and eventually ends in a domain structure state (the lowest frame in
Fig. \ref{denpos10}). The dotted line matches the dashed one in Fig. \ref{gplus}
and was obtained by solving the Euler--Lagrange equations corresponding to the
effective Lagrangian density (\ref{effective}) under the assumption of adiabatic 
approximation. Finally, the dotted--dashed line in Fig. \ref{gplus} represents
the local minimum with the densities of both fractions being shifted with respect
to each other (for large enough $g$ -- the middle frame in Fig. \ref{denpos10}). 
This minimum is obtained when at zero step of iteration one takes shifted left 
(for one component) and right (for another component) harmonic oscillator 
eigenstates (inset (b) in Fig. \ref{gplus}). 
Nothing changes qualitatively when one goes to larger systems as
shown in Fig. \ref{denpos50} which is the case of $50 + 50$ atoms system.
The dense structure visible in the lowest frame of Fig. \ref{denpos50} indicates
that for much larger number of atoms we should expect a smooth overlapping
distribution of both components in the corresponding case.

\section{Negative scattering length}
\label{minus}

\begin{figure}[htb]
\resizebox{2.7in}{2.1in}
{\includegraphics{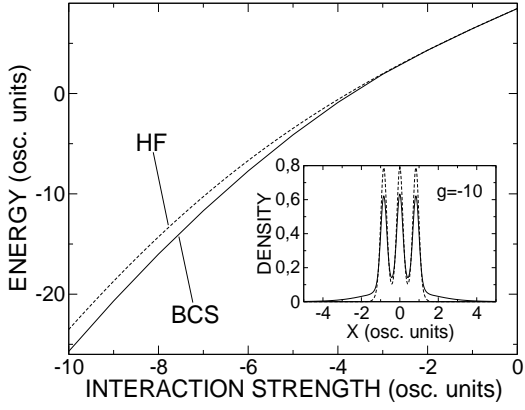}}
\caption{Ground state energy of $3 + 3$ atoms system as a function of 
interaction strength for both the Hartree--Fock and BCS approaches. For 
strong enough attraction the ground state of the system shows the appearance
of Cooper pairs. The inset shows the density (in oscillatory units) calculated 
via the Hartree--Fock (dashed line) and BCS (solid line) methods. Note that the 
densities for both fractions look the same.}
\label{enHF_BCS}
\end{figure}

\begin{figure}[htb]
\resizebox{2.5in}{4.4in}
{\includegraphics{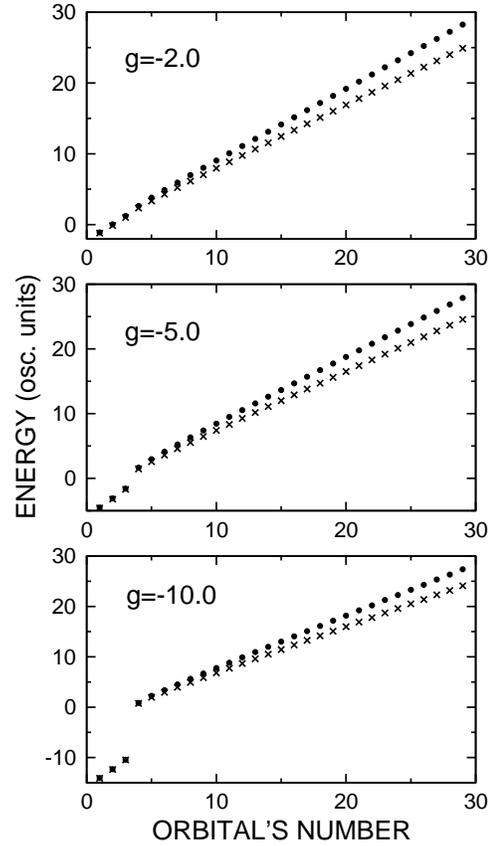}}
\caption{Single--particle Hartree--Fock energies for various strength
of attraction between atoms of different components. For strong enough 
attraction the energy gap is developed which separates off the populated 
levels. Dots and crosses correspond, respectively, to the first and second
component.}
\label{occupation}
\end{figure}

\begin{figure}[bht]
\resizebox{2.5in}{3.2in}
{\includegraphics{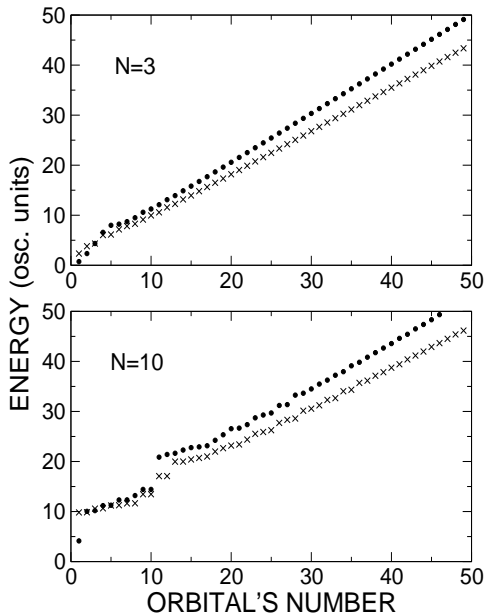}}
\caption{Single--particle Hartree--Fock energies for $g=+10.0$ and $3+3$
atoms system (the upper frame). No energy gap is developed. The lower frame
shows the same spectrum but in the case of $10+10$ atoms system in the
excited state with antiferromagnetic phase (as in the lowest part in Fig. 
\ref{denpos10}). Surprisingly, the energy gap is visible again although
there are some states within it.}
\label{occupation1}
\end{figure}

In this Section we consider the ground state of two--component system 
with attractive forces between atoms of different components. First of 
all, we have checked that in this case the Hartree--Fock energies and 
densities of the ground state do not depend on the choice of zero--step 
orbitals. The resulting ,,ground state'' energy is plotted as a dashed line 
in Fig. \ref{enHF_BCS} and the densities of both fractions (for $g=-10$) are 
shown in the inset (also by dashed line). It is clear that for strong enough 
attraction between atoms of different fractions a new density pattern with 
visible fragmentation is developed. Further analysis reveals the appearance 
of energy gap in the single--particle Hartree--Fock spectrum (see Fig. 
\ref{occupation}). The energy gap separates off the populated Hartree--Fock 
levels. If one of the atoms occupying the highest level is promoted to the 
state just above the energy gap the total energy of the system is increased 
and can be calculated based on formula (\ref{totenergy}). It turns out that the 
surplus in energy is of the order of the energy gap. Existence of such an 
energy gap in a single--particle spectrum could be a signature of appearance 
of pairing phenomenon in a system. This possibility will be verified in the 
following subsection. At the same time no such gap is developed when 
distinguishable atoms repel each other (Fig. \ref{occupation1}, the upper frame). 
Note, however, that an energy gap is also developed (Fig. \ref{occupation1}, 
the lower frame) in the case of repulsion between different atoms in 
antiferromagnetic phase present in excited states ( as in Fig. \ref{denpos10},
the lowest frame). 

\subsection{BCS formulation}

To check the idea whether the energy gap developed in a single--particle
energy spectrum is connected with the appearance of pairing phenomenon
in a system, we have reformulated our problem. Within Hartree--Fock approach
each particle interacts with all the others in the second component.
Now, we allow for formation of pairs including atoms of different spins.
More precisely, only configurations in which different atoms populate 
single--particle states of the same quantum number $n$ are permitted.
Hence, the ground state wave function is written in the following form
\cite{nucBCS}
\begin{equation}
|\Psi_{\rm BCS}> = \prod_n \, (u_n + v_n a_n^+ b_n^+ ) \, |\,0>  \, ,
\label{BCS}
\end{equation}
where $u_n^2 + v_n^2 = 1$ for each $n$ (this assures the normalization
condition) and operators $a_n^+,b_n^+$ create fermions, respectively, 
of first $(a_n^+)$ and second $(b_n^+)$ component. Of course, $v_n^2$
is the probability that the pair state $\varphi_n^{(1)} \varphi_n^{(2)}$
is occupied.

The Hamiltonian of the system includes kinetic and potential energies of 
both parts as well as the interaction between distinguishable atoms and is 
given by
\begin{eqnarray}
&&{\rm H} = \sum_{n,k} <n| E_{kin}^{(1)} + V_{trap}^{(1)} |\,k> a_n^+ a_k^{}     
\nonumber \\
&&+ \sum_{m,l} <m| E_{kin}^{(2)} + V_{trap}^{(2)} |\,l> b_m^+ b_l^{}
\nonumber \\
&&+ \sum_{n,m,k,l} <n\, m| V_{int} |\,k\, l> a_n^+ b_m^+ a_k^{} b_l^{}  \, .
\label{HAMIL}
\end{eqnarray}
The mean value of the Hamiltonian in the ground state given by (\ref{BCS}) is 
calculated as
\begin{eqnarray}
&&{\rm E}_{\rm BCS} = \sum_n (e_n^{(1)} + e_n^{(2)}) v_n^2 +
\sum_n V_{nn,nn} v_n^2  \nonumber \\
&&+ \sum_{\footnotesize \begin{array}{c} n,m \\ n\neq m \end{array}} 
(V_{nn,mm} u_n u_m v_n v_m  +  V_{nm,nm} v_n^2 v_m^2)  \, ,
\nonumber \\
\label{EBCS}
\end{eqnarray}
where
\begin{eqnarray}
&&e_n^{(1)} =\: <n| E_{kin}^{(1)} + V_{trap}^{(1)} |\,n>  \nonumber \\
&&e_n^{(2)} =\: <n| E_{kin}^{(2)} + V_{trap}^{(2)} |\,n>  \nonumber \\
&&V_{nn,mm} =\: <n\, n| V_{int} |\,m\, m>   \nonumber \\
&&V_{nm,nm} =\: <n\, m| V_{int} |\,n\, m>  \, .
\end{eqnarray}
The first term in (\ref{EBCS}) is the sum of kinetic and potential energies
of all atoms, the second and third (the first in the lower line sum) ones
represent the interaction energy between atoms within pairs whereas the last 
term describes the pairs interaction (i.e., the interaction between 
distinguishable atoms from different pairs).

Next, we minimize the expression (\ref{EBCS}) with respect to the real
amplitudes $u_n, v_n$ under extra condition that the total number of pairs
equals the number of atoms in each component $\sum_n v_n^2 = N$. To this
end, it is convenient to introduce new variables $\Theta_n$ in the following
way:  $u_n = \sin \Theta_n$ and $v_n = \cos \Theta_n$. Now, the energy of
the ground state ${\rm E}_{\rm BCS}$ is a function of angles $\Theta_n$
and its conditional minimum can be found by using the Lagrange multiplier
technique. One introduces the many--variable function 
${\rm F}_{\rm BCS} = {\rm E}_{\rm BCS} - \lambda\, G$ where 
$G = \sum_n \cos^2 \Theta_n  - N$  and looks for the unconditional minimum 
of function ${\rm F}_{\rm BCS}$. The necessary condition for that turns,
together with the normalization requirement for the total number of
pairs, to the following set of equations
\begin{eqnarray}
&&\sum_{m\neq n} V_{nn,mm} \sin{2 \Theta_m} - 2 \tan{2 \Theta_n}
\sum_{m\neq n} V_{nm,nm} \cos^2{\!\Theta_m}  \nonumber \\
&& = (e_n^{(1)} + e_n^{(2)} + V_{nn,nn} - \lambda) \tan{2 \Theta_n} 
\nonumber \\
&&n=0,1,2,...    \nonumber \\ 
&& \nonumber \\ 
&& \sum_m \cos^2{\!\Theta_m} = N  
\label{equBCS}
\end{eqnarray}
for variables $\Theta_n$ and the Lagrange multiplier $\lambda$. We solve 
numerically this set of equations and then check whether what we found 
is a minimum or not. We just randomly disturb obtained amplitudes 
$u_n, v_n$ and verify that the energy of the system according to
expression (\ref{EBCS}) always gets higher. 

\begin{figure}[bht]
\resizebox{2.7in}{2.1in}
{\includegraphics{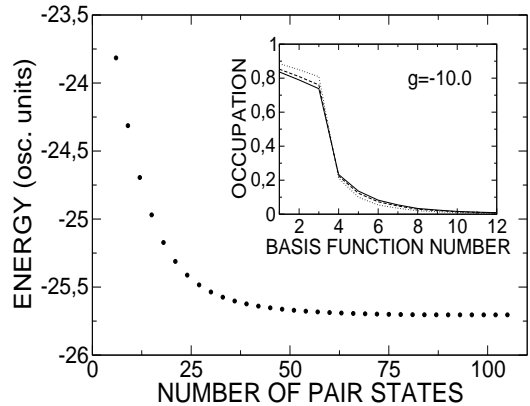}}
\caption{BCS energy of $3 + 3$ atoms system as a function of number of
states in the basis of pair states for $g=-10$. In this case $\rm{E}_{HF}=-23.4$
whereas $\rm{E}_{BCS}=-25.7$ osc. units. The inset shows corresponding 
occupations in cases when the number of basis states equals to $12$ (dotted 
line), $21$ (dashed line), and $105$ (solid line).}
\label{en_BCS}
\end{figure}

There is an important issue of choosing the appropriate two--particle basis 
functions for pair states. It turns out that it is good to build the pair
wave functions using single--particle Hartree--Fock orbitals. In such a case
the ground state of the system involves relatively small number of pair
states. In Fig. \ref{enHF_BCS} we plot the energy of the system of $3+3$ 
atoms as a function of the coupling constant $g$, calculated within both the 
Hartree--Fock and BCS approaches. It is seen that for strong enough attraction 
$(g<-3)$ the ground state of the system becomes of BCS type. Simultaneously,
the energy gap in a single--particle Hartree--Fock spectrum is being developed
(see Fig. \ref{denpos}). It is favorable for the system to lower its energy by 
forming pairs of atoms of different components. Fig. \ref{en_BCS} shows that, 
in fact, $50$ states basis is already good enough. The densities for both
fractions are plotted in the inset in Fig. \ref{enHF_BCS} for $g=-10$. In
BCS case the broad background is developed, not present for the Hartree--Fock
densities. In both cases the fragmentation pattern is visible which means that
one component serves the other as a periodic potential. This resembles somewhat
the situation when the atoms are confined in the optical lattice. For the 
interaction strength $g>-3$ one has $v_n\approx 1, u_n\approx 0\; {\rm for}\; n<3$ 
(and simultaneously $v_n\approx 0, u_n\approx 1\; {\rm for}\; n>3$) and there 
is no difference between the Hartree--Fock and BCS densities.

\begin{figure}[bht]
\resizebox{2.7in}{2.5in}
{\includegraphics{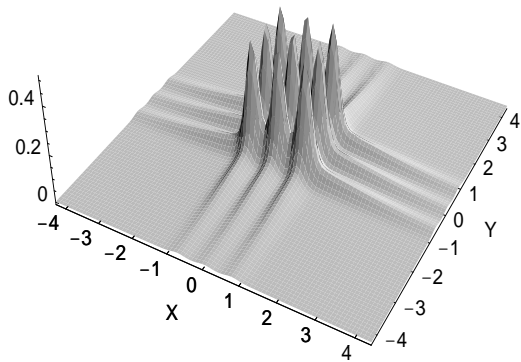}}
\caption{Diagonal part of two--particle density matrix for the
BCS ground state of $3+3$ atoms system, i.e., joined probability 
of detecting two distinguishable atoms. The interaction strength 
$g=-10$. All quantities are given in oscillatory units.}
\label{dmBCS}
\end{figure}

According to definition (\ref{mg1}) one can calculate the one--particle
density matrices for the BCS ground state given by (\ref{BCS})
\begin{eqnarray}
&&\rho_1^{(1)} ({\bf x}_1;{\bf x}'_{\!1}) =
\sum_{n} v_n^2\: \varphi_n^{(1)*}({\bf x}_1) \varphi_n^{(1)}({\bf x}'_{\!1})
\nonumber  \\
&&\rho_1^{(2)} ({\bf y}_1;{\bf y}'_{\!1}) =
\sum_{n} v_n^2\: \varphi_n^{(2)*}({\bf y}_1)  \varphi_n^{(2)}({\bf y}'_{\!1})
\end{eqnarray}
Assuming the energetically lowest $N$ pair states are fully occupied, i.e. 
$v_{n<N}=1,u_{n<N}=0 \;{\rm and}\; v_{n\geq N}=0,u_{n\geq N}=1$, the above 
formulas match the ones derived within the Hartree--Fock approach. The same is 
true also for the two--particle density matrix which is given by
\begin{eqnarray}
&&\rho_2^{(1)(2)} ({\bf x}_1,{\bf y}_1;{\bf x}'_{\!1},{\bf y}'_{\!1}) = 
\sum_{n,m}\; (u_n u_m v_n v_m + v_n^2 v_m^2) \times
\nonumber   \\\     
&&\varphi_n^{(1)*}({\bf x}_1)\,
\varphi_m^{(2)*}({\bf y}_1)\, \varphi_n^{(1)}({\bf x}'_{\!1})\,
\varphi_m^{(2)}({\bf y}'_{\!1})
\end{eqnarray}
In Fig. \ref{dmBCS} we have shown the diagonal part of two--particle density 
matrix for the BCS ground state of $3+3$ atoms system for the interaction
strength $g=-10$. For this value of $g$ the pairing phenomenon is already
well established and leads to $10\%$ decrease in energy in comparison with
the Hartree--Fock case. The characteristic crests along 'x' and 'y' axis
are developed as a result of presence of higher $n$ (bigger size understood
as the mean value of $|{\bf x}-{\bf y}|$) pair states in the BCS ground state
(see also Fig. \ref{enHF_BCS}). No such crests are visible for the Hartree--Fock 
,,ground state''.

\begin{figure}[bht]
\resizebox{2.7in}{2.5in}
{\includegraphics{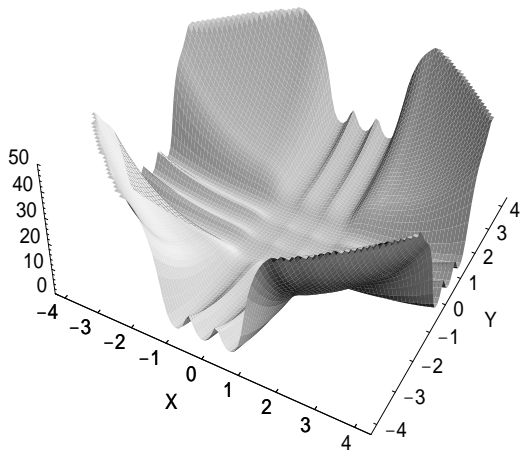}}
\caption{Correlation function for the BCS ground state of $3+3$
atoms system. The interaction strength $g=-10$. Only this part of 'xy'
plain is considered where the two--particle density matrix essentially
differs from zero. All quantities are given in oscillatory units.}
\label{cor3}
\end{figure}

\begin{figure}[thb]
\resizebox{2.5in}{3.2in}
{\includegraphics{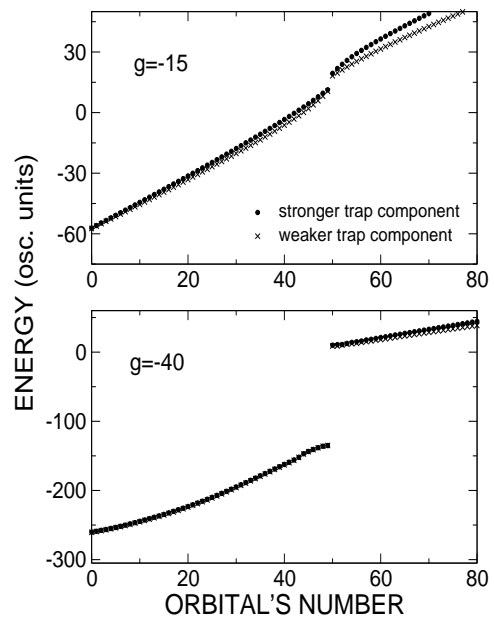}}
\caption{Single--particle Hartree--Fock energies for various strengths
of attraction between atoms of different components for a system of
$50+50$ atoms. For strong enough attraction the energy gap is developed
(upper frame) and the BCS ground state turns out to be reachable. When
the interaction strength is increasing further the energy gap is getting 
bigger (lower frame). }
\label{gap50}
\end{figure}

In Fig. \ref{cor3} we plot the correlation function $g_2$, defined generally as
\begin{eqnarray}
g_2({\bf x},{\bf y})=\frac{
\rho_2^{(1)(2)} ({\bf x},{\bf y};{\bf x},{\bf y}) -
\rho_1^{(1)} ({\bf x};{\bf x}) \rho_1^{(2)} ({\bf y};{\bf y})  }
{\rho_1^{(1)} ({\bf x};{\bf x}) \rho_1^{(2)} ({\bf y};{\bf y}) }  \;,
\nonumber \\
\end{eqnarray}
in the case of one--dimensional $3+3$ atoms system under the same parameters
as in the case of Fig. \ref{dmBCS}. It is true that the correlation function 
differs from zero mainly along the directions $y=x$ and $y=-x$. Whereas the 
former is not surprising because densities of both fractions look the same 
(see inset in Fig. \ref{enHF_BCS}), the latter means that the strong 
correlations are developed for distinguishable atoms being at positions $(2,4)$ 
and $(-4,-2)$ in oscillatory units. Since it corresponds to the size of pair 
states contributing essentially to the ground state of $3+3$ atoms system 
(approximately first 50 pair states -- see Fig. \ref{en_BCS}) it means that 
the presence of $y=-x$ correlations supports the explanation of properties 
of the ground state with the help of pairing phenomenon.

\begin{figure}[thb]
\resizebox{2.7in}{2.1in}
{\includegraphics{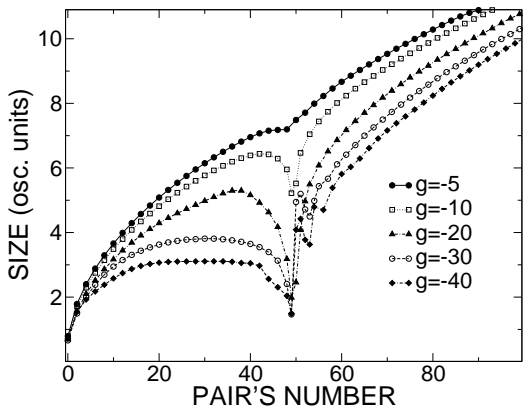}}
\caption{Size of the pair states calculated as the mean value of 
$|{\bf x}-{\bf y}|$ for the $50+50$ atoms system. Note that for strong
enough attraction the size is getting frozen and characteristic plateau is 
developed.}
\label{size}
\end{figure}

\begin{figure}[thb]
\resizebox{2.7in}{2.1in}
{\includegraphics{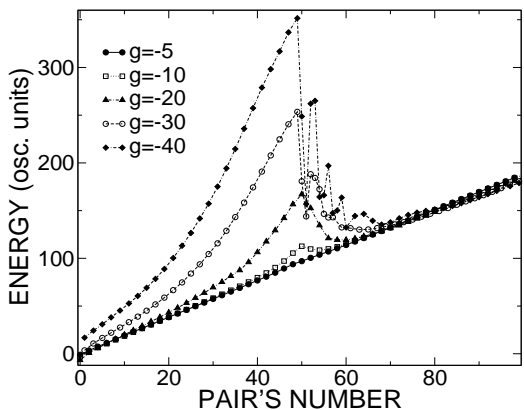}}
\caption{The energy of the pair states for the system of $50+50$ atoms.
For strong enough attraction the gap in the pairs spectrum is developed.}
\label{energy50}
\end{figure}

Increasing the number of atoms in both fractions we find further qualitative
changes in the properties of the ground state in comparison with $3+3$ case. 
When the coupling constant $g$ becomes smaller (i.e., for stronger attraction)
the energy gap in the Hartree--Fock spectrum gets bigger (Fig. \ref{gap50})
just like in the case of $3+3$ system. However, some new features appear
when one looks at the size of the pairs (Fig. \ref{size}) as well as at their 
energies (Fig. \ref{energy50}) for various values of the coupling constant $g$.
For strong enough attraction the size of the pairs is getting frozen revealing 
plateau and the gap is developed in the energy spectrum. The energy gap
in the pairs spectrum is perhaps the signature that the system enters
a new phase characterized by the appearance of clusters of particles
in its ground state, each built of four atoms.

It remains to be checked what is the range of validity of parameters 
used in our calculations. To this end, we show the following simple
estimation. Binary collisions dominate provided the mean interparticle
separation is much bigger than the s--wave scattering length, i.e.,
$n_{3D} |a_{3D}|^3 \ll 1$. According to Olshanii's result \cite{Olshanii},
for elongated systems the effective one--dimensional interaction strength
can be introduced via the relation
\begin{eqnarray}
g_{1D} = \frac{g_{3D}}{\pi a^2_{\bot}} \left( 1 - C\, \frac{a_{3D}}{a_{\bot}}
  \right)^{-1}  \;,
\label{g1D_Olshani}
\end{eqnarray}
where $g_{3D}=4\pi\hbar^2a_{3D}/m$, $C=1.4603$, and 
$a_{\bot}=(\hbar/m\omega_{\bot})^{1/2}$ determines the radial length scale.
Assuming the tight confinement one can introduce one--dimensional density
in such a way that
\begin{eqnarray}
n_{3D} = \frac{n_{1D}}{\pi a^2_{\bot}}  \;.
\end{eqnarray}
The range of validity of our model can be then expressed as
\begin{eqnarray}
\frac{n_{1D}|g_{1D}|^3 \alpha^2}
{\pi |g_{1D} C \alpha^{1/2} + 4|^3}  \ll 1   \;,
\label{validity}
\end{eqnarray}
where $\alpha = \omega_z / \omega_{\bot}$ and quantities $n_{1D}$ and
$g_{1D}$ are already taken in units related to one--dimensional system.
For example, for $|g_{1D}|=10$
the condition (\ref{validity}) leads to the relation $\alpha \ll 0.075$ 
(or $\omega_{\bot} / \omega_z \gg 13$) which, first, acts in the same direction 
as the assumptions leading to the formula (\ref{g1D_Olshani}) and, second,
can be fulfilled experimentally although for larger $|g_{1D}|$ the radial 
squeezing is getting stronger and stronger.

\section{Conclusions}
\label{concl}
We have analyzed the system of two--component Fermi gas confined in a
magnetic trap at zero temperature under assumption that the only 
interparticle interaction left is the repulsion or attraction between
atoms of different components. We employ the microscopic description of
the system by using the explicit atomic wave functions. The structure of 
the ground state strongly depends on the sign of the coupling constant. 
When distinguishable atoms repel each other strongly enough both fractions 
separate with that which is confined stronger filling the center of the trap. 
In the opposite case, i.e., when the atoms strongly attract each other, the 
ground state dramatically changes its character. The BCS pairs appear which 
is proved by showing that the energy of the system of pairs is lower than 
the corresponding Hartree--Fock energy and by analyzing the two--particle 
correlation function. For even stronger attraction the system enters a new 
phase where presumably clusters of particles are formed. We are working 
currently on extension of these calculations to the three--dimensional 
case as well as on inclusion of the optical lattice and finite temperatures.

\acknowledgments
M.B. acknowledges support by the Polish KBN Grant No. 2 P03B 052 24.
T.K. and K.R. were supported by the Polish Ministry of Scientific
Research Grant Quantum Information and Quantum Engineering
No. PBZ--MIN--008/P03/2003.

\end{document}